# Three-dimensional volumetric deconvolution in coherent optics and holography

Tatiana Latychevskaia

Physics Department, University of Zurich

Winterthurerstrasse 190, 8057 Zurich, Switzerland

tatiana@physik.uzh.ch

## Contents

# Abstract

Methods of three-dimensional deconvolution (3DD) or volumetric deconvolution of optical complex-valued wavefronts diffracted by 3D samples with the 3D point spread function are presented. Particularly, the quantitative correctness of the recovered 3D sample distributions is addressed. Samples consisting of point-like objects can be retrieved from their 3D diffracted wavefronts with non-iterative (Wiener filter) 3DD. Continuous extended samples, including complex-valued (phase) samples, can be retrieved with iterative (Gold and Richardson-Lucy) 3DD algorithms. It is shown that quantitatively correct 3D sample distribution can be only recovered with iterative 3DD, and with the optimal protocols provided. It is demonstrated that 3DD can improve the lateral resolution to the resolution limit and the axial resolution can be at least four times better than the resolution limit. The presented 3DD methods of complex-valued optical fields can be applied for 3D optical imaging and holography.

# 1. Introduction

Deconvolution with the point spread function (PSF) of an optical system has been applied in optical microscopy to improve the quality and resolution of the obtained images [1-4]. Three-dimensional deconvolution (3DD) or volumetric deconvolution of optical complex-valued wavefronts with the three-dimensional (3D) PSF was demonstrated in [5], where two types of 3DD were presented, namely, non-iterative (Wiener filter) and iterative deconvolution. The recovered 3D sample distributions demonstrated an enhanced lateral and axial resolution and no out-of-focus signal. The present study explores 3DD algorithms which allow for a quantitatively correct recovery of 3D sample distributions and quantitatively estimates the resolution enhancement.

    In general, the 3D Fourier spectrum of the output signal (image) can be represented as a product of the 3D Fourier spectrum of the input signal (sample) and the 3D Fourier spectrum of the

PSF or the optical transfer function (OTF). Thus, the OTF describes what frequencies of the signal are transferred to the output image formed by the optical system. The OTF typically has only a limited range of non-zero values, meaning that all other frequencies of the input signal are multiplied with zeroes and therefore do not contribute to the formation of the output image. In order to restore the input signal from the output signal, these missing frequencies need to be recovered. Thus, 3DD ideally should restore the missing frequencies in the 3D Fourier spectrum of the input signal, and with this, the distribution of the input signal.

To acquire all the sample frequencies, the 3D OTF distribution can be used as a tool to "scan" over the 3D Fourier spectrum of the sample, which can be realized, for example, by a tomography approach [6]. 3DD can enhance the z resolution of reconstructed objects by improving the depth of focus and the lateral resolution of the imaged samples [7, 8]. In this study, we consider an optical system that does not employ a complicated setup, such as tomography, and where the scattered wavefront is measured in one plane (single-shot measurement). In addition, we will consider 3DD performed on complex-valued wavefronts and optical wavefronts reconstructed from in-line holograms [9]. The complex-valued wavefronts, such as wavefront diffracted by a sample, can be directly measured by employing an interferometric technique, as for example, off-axis [10-12] or phase shifting [13] holography.

## 2. 3D volumetric deconvolution in coherent optics
### 2.1 Principles of image formation by convolution

The image formed by the optical system $u_{\text{out}}(\vec{r})$ can be represented as a convolution of the input wavefront $u_{\text{in}}(\vec{r})$ with the PSF of the system $\tau(\vec{r})$:

$$u_{\text{out}}(\vec{r}) = u_{\text{in}}(\vec{r}) \otimes \tau(\vec{r}), \qquad (1)$$

where the PSF is the output signal to the input signal in the form of a point source or a point-like object: $u_{\text{out}}(\vec{r}) = \delta(\vec{r}) \otimes \tau(\vec{r}) = \tau(\vec{r})$, where $\vec{r} = (x, y, z)$ is the coordinate and $\otimes$ denotes the convolution. In general, the PSF is a 3D complex-valued function. Equation (1) can be rewritten using the convolution theorem: $\text{FT}(u_{\text{out}}) = \text{FT}(u_{\text{in}})\text{T}$, where $\text{T} = \text{FT}(\tau)$ is the Fourier transform (FT) of the PSF, which is the optical transfer function (OTF) of the optical system.

By neglecting multiple scattering, diffraction on a sample can be described in similar terms. The sample can be considered as the input signal and the diffracted wave as the output signal. The diffracted wave can be represented as a convolution of the product of the sample function $s(\vec{r})$ with the incident wave $u_0(\vec{r})$ and the wave diffracted on a point-like object $p(\vec{r})$:

$$u(\vec{r}) = [u_0(\vec{r})s(\vec{r})] \otimes p(\vec{r}). \qquad (2)$$

This statement agrees well with the Huygens-Fresnel principle that each point is a source of secondary waves:

$$u(\vec{r}) = \iiint u_0(\vec{\rho})s(\vec{\rho}) \frac{\exp(ik|\vec{\rho}-\vec{r}|)}{|\vec{\rho}-\vec{r}|} d\vec{\rho},$$

where $k = \frac{2\pi}{\lambda}$ is the wavenumber, if we set

$$p(\vec{r}) = \frac{\exp(ikr)}{r}. \qquad (3)$$

## 2.2 Analytical wavefront from a point-like source

$p(\vec{r})$ given by Eq. (3) is a spherical wave originating from a point source. The 3D FT of $p(\vec{r})$ can be calculated analytically and gives:

$$U(\alpha,\beta,\gamma) = \iiint \frac{\exp(ikr)}{r} \exp\left[-\frac{2\pi i}{\lambda}(x\alpha + y\beta + z\gamma)\right] dxdydz =$$
$$= 4\pi\left[i\pi\delta'(\varsigma-k) - \frac{1}{(\varsigma-k)^2}\right],$$

where $\vec{\varsigma} = (\alpha,\beta,\gamma)$ is the coordinate in the Fourier domain and $\delta'(\varsigma)$ is the generalized derivative of the Dirac delta impulse. The amplitude of $U(\alpha,\beta,\gamma)$ exhibits a distribution that has non-zero values on a sphere $\varsigma = k$. Different wavelengths result in different radii of the sphere of non-zero values in Fourier space. Figures 1(a)-(i) show the 3D distributions of $p(\vec{r})$ and its 3D FT (calculated as the 3D fast Fourier transform (FFT)) simulated with wavelength $\lambda$ = 500 nm, an *x, y, z* range of *s* = 10 µm and sampling with 200 × 200 × 200 pixels. Figures 1(j)-(r) shows that when the *z* range is larger than the *x, y* range (here the *z* range is *s* = 20 µm), the sphere of non-zero values in Fourier space turns into an ellipsoid-like surface. For z range much larger x,y range, the sphere of non-zero values turns into a parabola-like distribution, as shown in the next section.

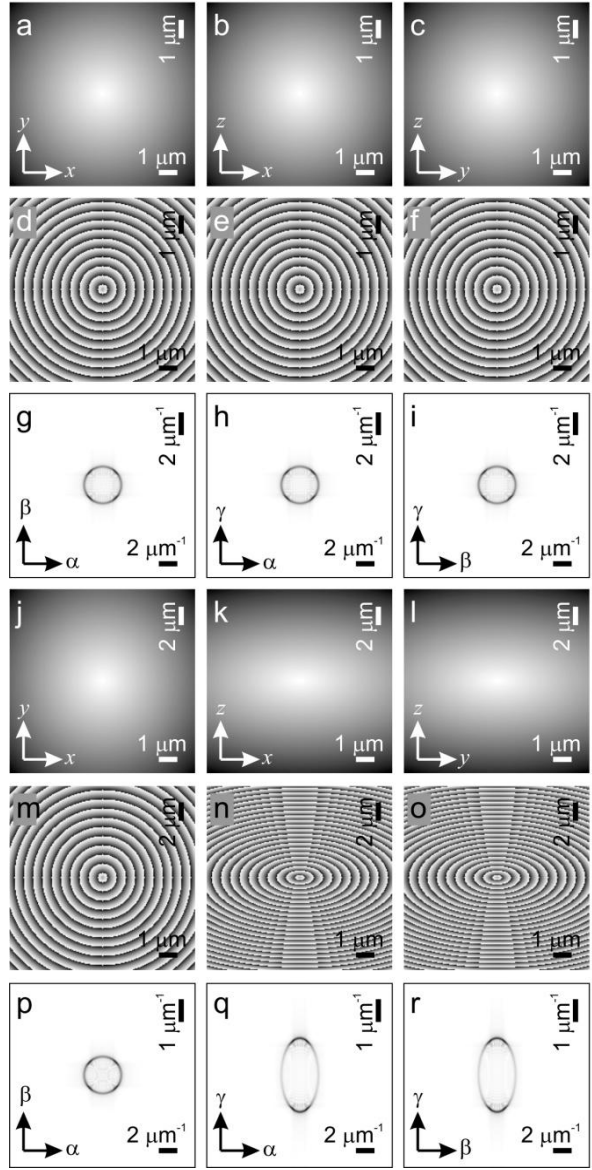

Fig. 1. 3D distribution of $p(\vec{r}) = \dfrac{\exp(ikr)}{r}$ and its 3D FT, for $z$-range of (a)-(i) 10 μm and (j)-(r) 20 μm. Distributions in the $(x, y, 0)$, $(x, 0, z)$ and $(0, y, z)$ planes are shown. (a)–(c), (j)-(l) Amplitude and (d)–(f), (m)-(o) phase distributions of $p(\vec{r})$. (g)–(i), (p)-(r) Distributions of the amplitude of 3D FT of $p(\vec{r})$.

## 2.3 Simulated wavefront from a point-like source

In optics, the $z$ range is typically much larger than the $x$, $y$ range and a paraxial approximation can be applied. The wavefront originating from a point-like source $\delta(x, y)$ can be calculated by the angular spectrum method (ASM) [14-16] using the following steps:

(i) The 2D FT of a point-like source is calculated and gives:

$$P_0(\alpha,\beta) = \iint \delta(x,y)\exp\left[-\frac{2\pi i}{\lambda}(x\alpha+y\beta)\right]dxdy = 1; \quad (4)$$

(ii) The result of (i) is multiplied by the function:

$$\exp\left(\frac{2\pi i}{\lambda}z\sqrt{1-\alpha^2-\beta^2}\right); \quad (5)$$

(iii) The inverse 2D FT of (ii) is calculated as:

$$p(x,y,z) = \iint \exp\left(\frac{2\pi i}{\lambda}z\sqrt{1-\alpha^2-\beta^2}\right)\exp\left[\frac{2\pi i}{\lambda}(x\alpha+y\beta)\right]d\alpha d\beta, \quad (6)$$

and gives the complex-valued distribution of the wavefront in the $(x,y)$-plane located at distance $z$. By calculating the wavefront at different $z$ distances, a 3D complex-valued wavefront distribution $p(x,y,z)$ is obtained. The 3D FT of $p(x,y,z)$ can be calculated analytically:

$$P(\alpha,\beta,\gamma) = \iiint p(x,y,z)\exp\left[-\frac{2\pi i}{\lambda}(x\alpha+y\beta+z\gamma)\right]dxdydz = \delta\left(\gamma-\sqrt{1-\alpha^2-\beta^2}\right),$$

which is a distribution of non-zero values on a sphere. Figure 2 shows $p(x,y,z)$ and its 3D FT simulated with wavelength $\lambda$ = 500 nm, an *x*, *y* range of $s_{x,y}$ = 400 μm, a z range of $s_z$ = 2 mm and sampled with 200 × 200 × 200 pixels.

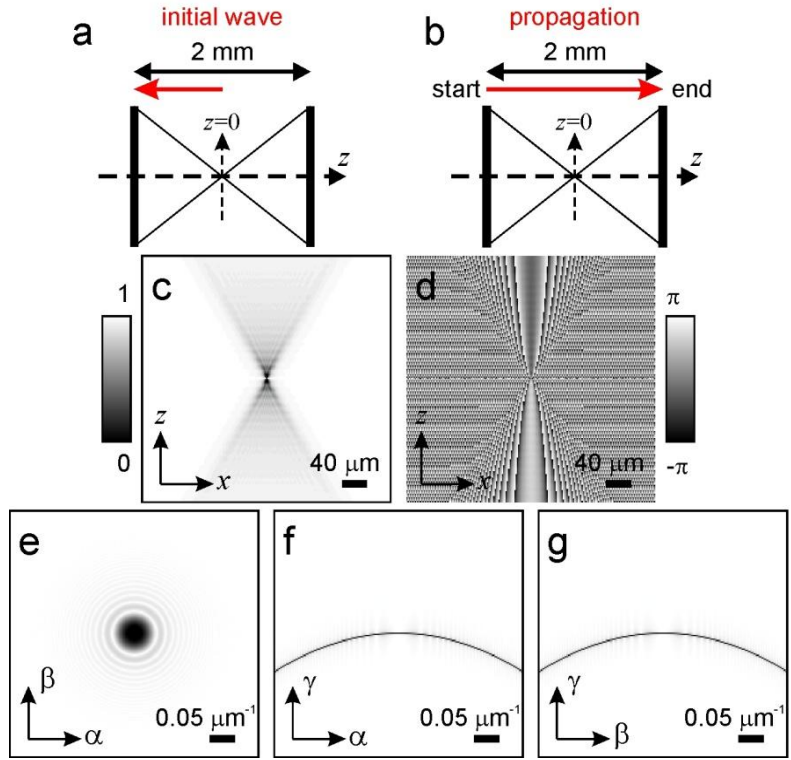

Fig. 2. 3D distribution of $p(x,y,z)$ calculated by Eqs. (4)–(6) and its 3D FT.
(a) The initial wavefront is obtained at a plane *z* = -1 mm and is then (b)

propagated along the positive direction of the *z* axis from *z* = -1 mm to *z* = 1 mm. (c) and (d) Distribution of amplitude and phase in the $(x,0,z)$ plane, respectively. (e)–(g) 2D distributions of the amplitude of $P(\alpha,\beta,\gamma)$, where $P(\alpha,\beta,\gamma)$ calculated as the 3D FFT of $p(x,y,z)$.

Note that in Fig. 2(d), the phase distribution of the convergent wave displays positive values for $z < 0$, which seem to contradict the rule that a convergent wave is described by $\exp(-ikr)$ but can be explained by the fact that the displayed phase is wrapped between 0 and 2π and therefore might exhibit negative or positive values. The phase of the wavefront originating from a point-like source is given by $kr$ and has the same sign for positive and negative $z$ (Fig. 1). In the case of the wavefront propagation, the phase of the wavefront changes its sign when the wavefront passes through the origin of the point-like source (Fig. 2(d)). Since wavefront propagation integrals are typically solved in some approximations, the phase of the spherical wave loses this *z* ambiguity, for example, $\exp(ikr) \approx \exp\left(\frac{2\pi i}{\lambda}z\right)\exp\left[\frac{i\pi}{\lambda z}\left(x^2+y^2\right)\right]$ in paraxial approximation.

## 3. Deconvolution algorithms
### 3.1 Non-iterative 3DD (Wiener filter)

The non-iterative 3DD (Wiener filter) can be applied when the sample consists of individual isolated scatterers, as, for example, particles in a particle flow. The 3DD is performed on intensities by the following transformation [5]:

$$s'(\vec{r}) = \mathrm{FT}^{-1}\left(\frac{\mathrm{FT}\left[|u(\vec{r})|^2\right]}{\mathrm{FT}\left[|p(\vec{r})|^2\right]+\beta_0}\right),$$

where $\beta_0$ is a small constant added to avoid division by zero. The obtained distribution $s'(\vec{r})$ is not the distribution of the sample but the distribution of the positions of the individual scatterers. The 3DD performed on intensities can successfully reconstruct the positions of the scatterers [5, 17], which can be explained by the broad distribution of $\mathrm{FT}\left[|p(\vec{r})|^2\right]$ when compared to $\mathrm{FT}\left[p(\vec{r})\right]$.

Both distributions are shown for comparison in Fig. 3. Here, the simulation is mimicking an experimental situation: a complex-valued wavefront is available at the detector plane and p(x,y,z) is obtained by propagating the wavefront backwards. The parameters of the simulation are provided in Appendix A. Non-iterative 3DD allows for very precise recovery of the positions of scatterers, which is practical when the positions (but not the shape) of the scatterers are important, as, for example, in particle tracking [17]. Note that different directions of the wavefront propagation (Figs. 2(a,b) and 3(a,b)) result in flipping of the spherical surface of non-zero values in Fourier space, as can be seen by comparing Figs. 2(f) and (g) and 3(d) and (e).

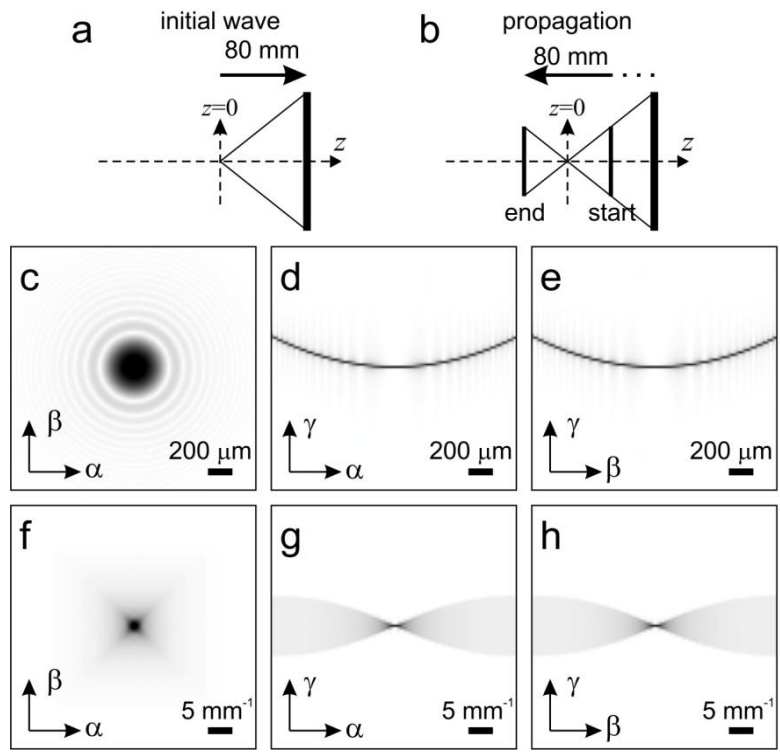

Fig. 3. 3D distribution of 3D FT $p(x, y, z)$ and $|p(x, y, z)|^2$, where $p(x, y, z)$ is calculated by Eqs. (4)–(6) and 3D FT is calculated by 3D FFT. (a) Initial wavefront obtained at a plane $z$ = 80 mm. (b) The obtained wavefront is then propagated along the negative direction of the $z$ axis from $z$ = 40 to -40 mm. (c)–(e) 2D distributions of the amplitude of 3D FT of $p(x, y, z)$. (f)–(h) 2D distributions of the amplitude of 3D FT of $|p(x, y, z)|^2$.

## 3.2 Iterative 3DD

Iterative 3DD recovers the sample distribution $s(\vec{r})$ from the scattered wavefront $u(\vec{r})$ by accurately solving Eq. (2) in an iterative manner. Iterative 3DD is performed on complex-valued wavefronts and has much a broader range of application, for example, for extended objects, phase objects and for all objects where non-iterative 3DD can be applied.

### 3.2.1 Gold algorithm

The Gold algorithm for iterative deconvolution includes the following steps [18]:

$$s^{(1)}(\vec{r}) = u(\vec{r})$$

(i) $\quad u^{(k)}(\vec{r}) = s^{(k)}(\vec{r}) \otimes p(\vec{r})$ (7)

(ii) $\quad s^{(k+1)}(\vec{r}) = s^{(k)}(\vec{r}) \dfrac{u(\vec{r})}{u^{(k)}(\vec{r})}$

(iii) $\quad k = k+1$

To avoid division by zero in step (ii), the division can be calculated as Wiener filter:

$$s^{(k+1)}(\vec{r}) = s^{(k)}(\vec{r}) \dfrac{u(\vec{r})\left[u^{(k)}(\vec{r})\right]^{*}}{\left|u^{(k)}(\vec{r})\right|^{2} + \beta_{0}} \quad (8)$$

where $\beta_0$ is a small addend that should be selected to be much smaller than the maximal value of $\left|u^{(k)}(\vec{r})\right|^2$ but sufficiently large to ensure convergence of the iterative process. Normalization of $u^{(k)}(\vec{r})$ can be carried out after step (i) to avoid division by very small values::

$$u^{(k)}(\vec{r}) = \dfrac{\left\{u^{(k)}(\vec{r}) - \min\left[u^{(k)}(\vec{r})\right]\right\}\left\{\max\left[u(\vec{r})\right] - \min\left[u(\vec{r})\right]\right\}}{\left\{\max\left[u^{(k)}(\vec{r})\right] - \min\left[u^{(k)}(\vec{r})\right]\right\}} + \min\left[u(\vec{r})\right]. \quad (9)$$

The normalization ensures that the normalized $u^{(k)}(\vec{r})$ has the same maximum and minimum as $u(\vec{r})$. With each iteration step, the distributions $s^{(k)}(\vec{r})$ and $u^{(k)}(\vec{r})$ approach the complex-valued distributions of the true 3D sample distribution $s(\vec{r})$ and the diffracted wave $u(\vec{r})$, respectively, while $p(\vec{r})$ remains unchanged. Gold's algorithm can quantitatively retrieve the correct distribution of signal $s(\vec{r})$.

### 3.2.2 Richardson-Lucy algorithm

The Richardson-Lucy (RL) deconvolution algorithm includes the following steps [19, 20]:

$$s^{(1)}(\vec{r}) = u(\vec{r})$$

(i) $\quad u^{(k)}(\vec{r}) = s^{(k)}(\vec{r}) \otimes p(\vec{r})$

(ii) $\quad s^{(k+1)}(\vec{r}) = s^{(k)}(\vec{r}) \left[\dfrac{u(\vec{r})}{u^{(k)}(\vec{r})} \otimes \widehat{p}(\vec{r})\right]$

(iii) $\quad k = k+1$

where $\widehat{p}(\vec{r})$ is a flipped $p(\vec{r})$. The RL algorithm allows for the quantitatively correct reconstruction of the sample distribution, provided $p(\vec{r})$ is normalized so that $\iiint p(\vec{r})\mathrm{d}\vec{r}=1$.

Our studies showed that when applied to optical fields, the RL algorithm quickly diverges and applying the Wiener division in step (ii) and/or adding normalization of $u^{(k)}(\vec{r})$ after step (i) do not help. The failure can be explained by the fact that $u^{(k)}(\vec{r})/u(\vec{r})$ in step (ii) should approach a uniform distribution of ones (a distribution where all values are one) as the iterative process progresses. However, at the first iteration, $u^{(k)}(\vec{r})/u(\vec{r})$ is far from the distribution of ones and additional convolution with $\widehat{p}(\vec{r})$ (this operation is not carried out in Gold's algorithm) makes is even less similar to the distribution of ones. This issue becomes more pronounced with each following iteration, eventually causing the whole iterative process to diverge. For this reason, we use only Gold's algorithm for iterative 3DD.

## 4. Sampling and resolution

Correct sampling of optical wavefronts is crucial for numerical calculations when FFTs are employed, as in the ASM. The phase distribution of the propagator in the ASM (Eq. (5)) is given by:

$$\frac{2\pi z}{\lambda}\sqrt{1-\alpha^2-\beta^2} \approx \frac{\pi z}{\lambda}\left(\alpha^2+\beta^2\right). \tag{10}$$

When digitally sampled, the coordinates are replaced by $\alpha=\Delta_\alpha l,\ \beta=\Delta_\beta m$, where $l,m=-N/2...N/2$ are the pixel numbers and the pixel size in the Fourier domain is given by $\Delta_\alpha=\Delta_\beta=\frac{\lambda}{s}$, where s × s is size of the area in the real space. Thus, we can rewrite Eq. (10) as $\frac{\pi z}{\lambda}\left(\alpha^2+\beta^2\right)=\frac{\pi z\lambda}{s^2}\left(l^2+m^2\right)$. According to the Nyquist-Shannon theorem [21, 22], a signal is correctly sampled when its highest frequency component is sampled with at least two pixels per period. The value of the phase at the first pixel (l = -N/2 and m = 0) is given by: $\varphi_1=\frac{\pi\lambda z}{s^2}\left(\frac{N}{2}\right)^2$ and the value of the phase at the second pixel is given by: $\varphi_2=\frac{\pi\lambda z}{s^2}\left(\frac{N}{2}-1\right)^2$. The difference between the two values is given by: $\varphi_1-\varphi_2=\frac{\pi\lambda z}{s^2}\left(\frac{N}{2}\right)^2-\frac{\pi\lambda z}{s^2}\left(\frac{N}{2}-1\right)^2\approx\frac{\pi\lambda z}{s^2}N$, where we assume that the number of pixels is sufficiently large N >>1. The phase difference should not exceed half a period, i.e., $\varphi_1-\varphi_2\leq\pi$, which gives the maximal allowed number of pixels for correct sampling: $N_{max} \leq \dfrac{s^2}{z\lambda}$. Alternatively, the minimal pixel size, which allows for correct sampling, is given by:

$$\Delta_{min} = \frac{s}{N_{max}} \geq \frac{z\lambda}{s}. \tag{11}$$

## 5. 3D volumetric deconvolution of optical wavefronts
### 5.1 Point-like sources and scatterers

We consider three points sources at different $(x, y, z)$ positions, with each source emitting a spherical wave $\exp(ikr)/r$ and its complex-valued distribution calculated in the detector plane by Eqs. (4)–(6), as sketched in Fig. 4(a). The parameters of the simulation are provided in Appendix A. A sum of the scattered waves is calculated in the detector plane and the obtained wavefront is propagated back, thus forming a 3D wavefront, as sketched in Fig. 4(b). The 2D distributions of the amplitude of the 3D wavefront in the planes $(x, y, 0)$, $(x, 0, z)$ and $(0, y, z)$ are shown in Fig. 4(c)–(e), with the out-of-focus signal evident in the shown distributions. The PSF is formed with the wavefront from the source at $(0,0,0)$ calculated in the detector plane, which is then propagated back and the obtained 3D wavefront is the PSF.

Deconvolution was performed by non-iterative 3DD and by iterative 3DD using Gold's algorithm with Wiener division (Eq.(8)) and normalization (Eq. (9)). The results are shown in Figs. 4(f)–(h) and (i)–(k), respectively. Both non-iterative and iterative 3DD deliver the correct results, with the three individual sources correctly spatially reconstructed. However, only iterative 3DD delivers quantitatively correct values of the source amplitudes.

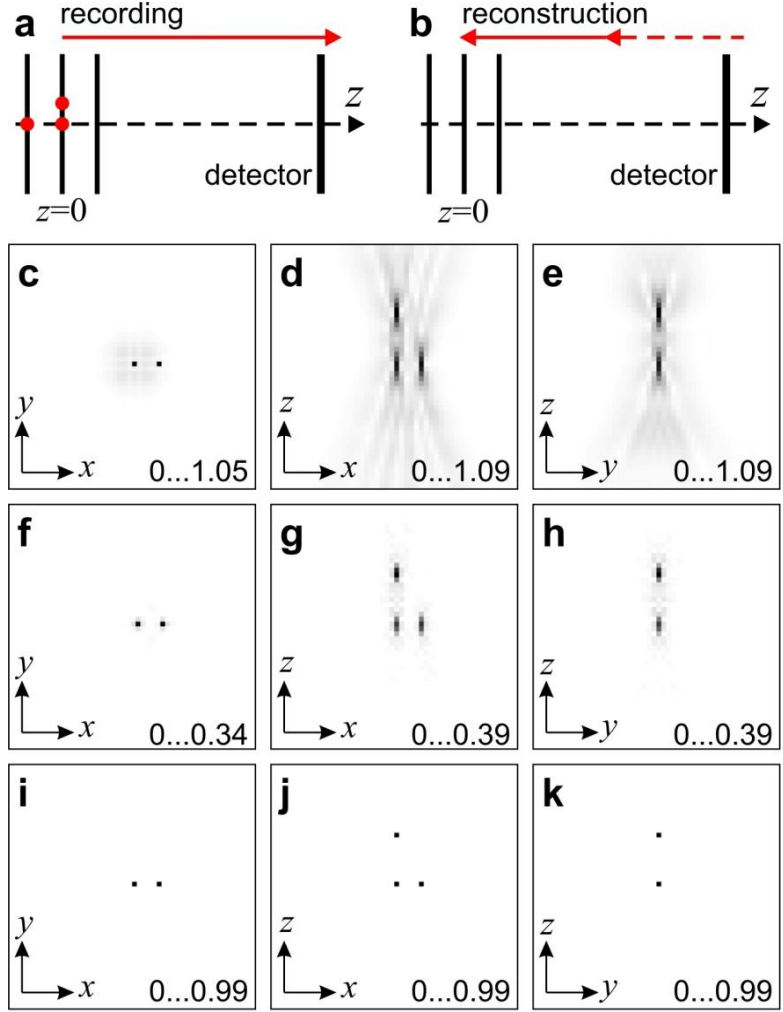

Fig. 4. 3DD of optical complex-valued wavefronts originating from three point sources. (a) and (b) Propagation of the wavefronts during recording and reconstruction. (c)–(e) 2D distributions of the amplitude of 3D wavefront obtained by back propagation from the detector plane. (f)–(h) Amplitude distribution obtained by 3D non-iterative (Wiener) deconvolution, $\beta_0$ = 0.01. (i)–(k) Amplitude distribution obtained after applying 3D iterative (Gold's algorithm) deconvolution for 10000 iterations, $\beta_0$ = 0.01. In (c)–(k), only the central 50 × 50 pixels regions are shown. The values given at the bottoms are the range of the amplitude values.

A similar situation occurs when three point-like scatterers are illuminated by a plane wave, in which case, each scattered wave has an additional factor $\exp(ikz)$. The results (not shown here) are very similar to the results for individual sources. Both deconvolution approaches deliver correct results, with three individual scatterers correctly located, but only iterative 3DD delivers quantitatively correct values of the source amplitude.

## 5.2 Continuous sample

A continuous sample cannot be represented as a set of isolated point-like scatterers and iterative 3DD must be applied. 3DD of optical wavefronts of a continuous sample in the form of three letters "α", "β" and "γ" was previously demonstrated in [5]. Here, we consider samples that are more extended, such as a patch, and investigate the quantitative correctness of the retrieved distributions. In the examples below, 3DD was performed by applying Gold's algorithm (Eq. (7)) using three different sets of filters:

**Filters1**: Low-pass and normalization filters are additionally applied. (a) After step (i) the factor is calculated: $f = \dfrac{\max |u(\vec{r})|}{\max |u^{(k)}(\vec{r})|}$. (b) Normalization is performed as described by Eq. (9). Without this normalization, the recovered 3D distributions appear not to be quantitatively correct. (c) The division in step (ii) is completed by applying a Wiener filter, as in Eq. (8). (d) At every fifth iteration, a low-pass filter ($D$ = 5 pixels) is applied to the 3D sample distribution $s^{(k+1)}(\vec{r})$. The distribution of the low-pass filter in the Fourier domain is calculated as:

$$\mathrm{LP} = \exp\left(-\frac{l^2 + m^2 + n^2}{2D^2}\right), \tag{12}$$

where $l$, $m$ and $n$ are the pixel indices, and $D$ is the standard deviation. (e) At the end of each iteration, the obtained sample distribution is normalized as: $s^{(k+1)}(\vec{r}) = s^{(k+1)}(\vec{r}) f$, where $f$ is calculated in (a).

**Filters2**: The same as Filters1 with additional loose mask object support. A mask support sets all values outside the mask to zero and keeps the values inside the mask unchanged.

**Filters3**: Simple division in step (ii) (no Wiener filter (Eq. (8)), no normalization (Eq. (9))) and a tight mask object support are applied.

### 5.2.1 Real-valued sample

A real-valued sample was chosen in the form of a square patch, as shown in Figs. 5(a)–(c). The patch size was 200 μm × 200 μm or 10 × 10 × 1 pixels. The parameters of the simulation are provided in Appendix A. The amplitude of the diffracted wavefront is shown in Figs. 5(d)–(f). The results of 3DD using Filters1 with $\beta_0$ = 0.01 after 999 iterations are shown in Figs. 5(g)–(i). Note that the retrieved thickness of the patch is slightly larger than 1 pixel (Figs. 5(h) and (i)). The results of 3DD using Filters2 with a loose mask support of 40 × 40 × 4 pixels, after 9999 iterations, are shown in Figs. 5(j)– (l). The results of 3DD using Filters3 with a tight mask object support of 11 × 11 × 1.6 pixels, after 100 iterations, are shown in Figs. 5(m)–(o).

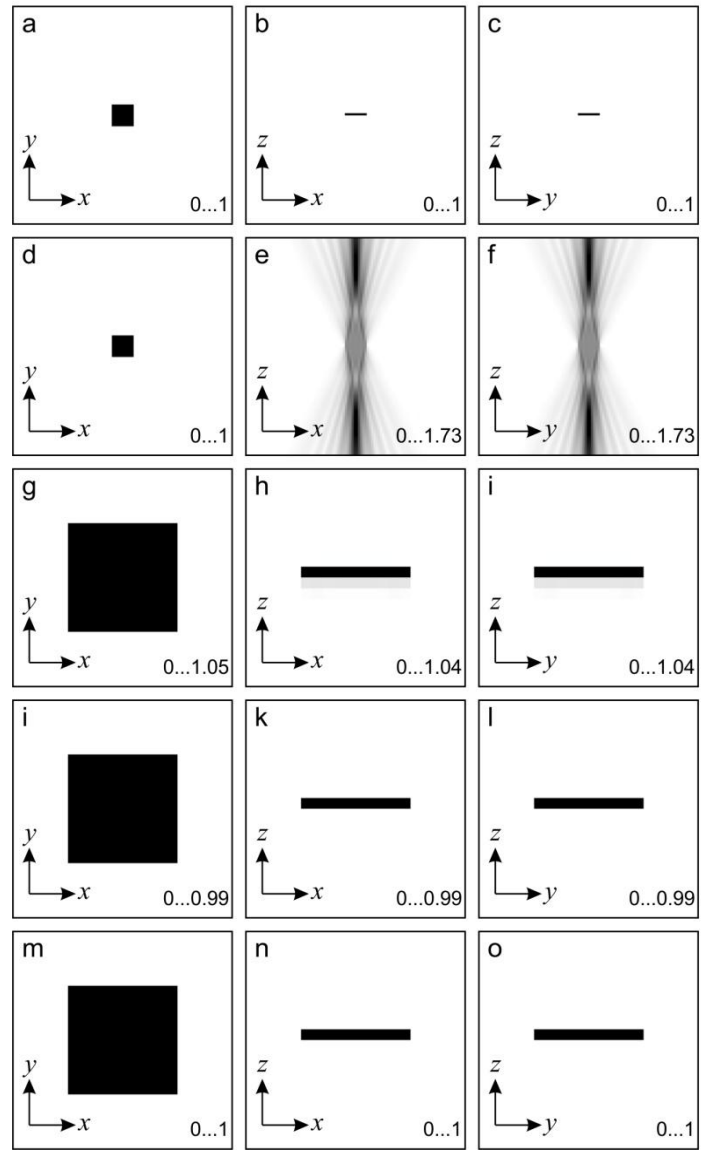

Fig. 5. 3DD applied to the 3D wavefront reconstructed from a hologram of a real-valued sample. 2D amplitude distributions in the $(x, y, 0)$, $(x, 0, z)$ and $(0, y, z)$ planes are shown. The contrast is inverted. (a)–(c) Distribution of the original sample. (d)–(f) Amplitude of the diffracted wavefront. (g)–(o) Amplitude of the sample distribution recovered after 3DD using filters: (g)–(i) Filters1, (j)–(l) Filters2 and (m)–(o) Filters3. In (g)–(i), only the central 20 × 20 pixel regions are shown. The values given at the bottom are the range of the amplitude values.

Quantitatively, all three filter sets show the correct results. Filters3 requires the least amount of iterations, but it also requires knowledge of the exact sample shape in order to create a tight mask object support. When the sample shape is unknown, Filters1 or Filters2 can be used to obtain an initial sample distribution and to create a tight mask object support. Filters1 or Filters2 require no a priori information regarding the sample.

### 5.2.2 Complex-valued (phase) sample

Another example is a complex-valued sample in the form of a square patch with an additional phase shift of 1 radian. The size of the patch and the parameters for simulation and reconstruction are the same as in the previous example, as shown in Figs. 6(a)–(c). The amplitude of the diffracted wavefront is shown in Figs. 6(d)–(f). The results obtained after 3DD using Filters2 with loose mask support of 40 × 40 × 4 pixels, after 9999 iterations, are shown in Figs. 6(g)–(l). The results obtained after 3DD using Filters3 with a tight mask object support of 11 × 11 × 1.6 pixels, after 100 iterations, are shown in Figs. 6(m)–(s). Here, also, both Filters2 and Filters3 retrieve the quantitatively correct amplitude and phase distributions of the sample. Filters3 requires the least iterations but also requires knowledge of the exact sample shape in order to create a tight mask object support. When the sample shape is not known, Filters2 can be used.

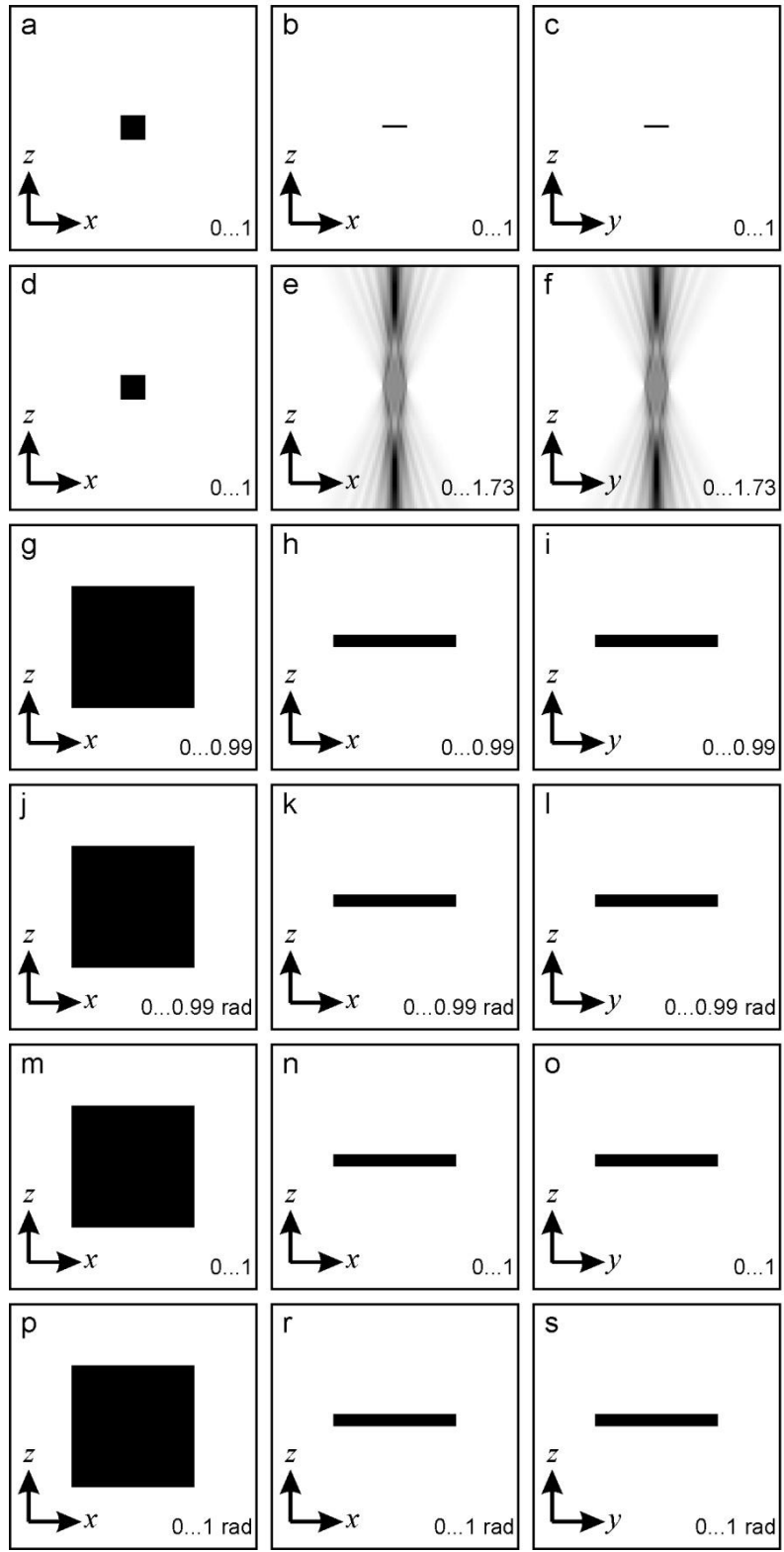

Fig. 6. 3DD applied to a 3D wavefront reconstructed from a hologram of a complex-valued sample. 2D distributions in the $(x, y, 0)$, $(x, 0, z)$ and $(0, y, z)$ planes are shown. The contrast is inverted. (a)–(c) Amplitude distribution of the sample, the phase distribution is the same with the maximal phase shift of 1 radian. (d)–(f) Amplitude of the diffracted wavefront. Amplitude and phase distributions, respectively, obtained after 3DD with (g)–(i) and (j)–(l) Filters2; (m)–(o) and (p)–(s) Filters3. In (g)–(s), only the central 20 × 20 pixel regions are shown. The values given at the bottom of (a) - (i) and (m) - (o) are the range of the amplitude values.

## 6. 3D deconvolution in holography

Holography allows for the restoration of a 3D sample distribution from a 2D holographic record [9, 23]. However, when a hologram is reconstructed, the result is not the sample distribution but the wavefront diffracted by the sample. The wavefront carries information regarding all the diffraction events that took place in the sample. The wavefront can be reconstructed at any plane from the hologram, thus giving rise to a 3D reconstructed wavefront. This wavefront is identical to the wavefronts discussed above and all the results and conclusions derived for 3DD of the wavefronts also apply here. In the case of off-axis holography, the exact complex-valued wave diffracted by the sample is reconstructed and 3DD can be applied as described above. In the case of in-line or Gabor-type holography [9, 23], besides the complex-valued wave scattered by the sample, the signal due to the twin image [23, 24] and object terms are also reconstructed. By applying 3DD, these artifactual signals can be successfully eliminated and the 3D sample distribution can be restored [5]. The proposed in [5] non-iterative 3DD has been successfully applied for 3D particle field reconstruction and particle tracking [17, 25, 26]. Iterative 3DD was recently applied for restoring the 3D distribution of biological samples [27]. 3DD in holography can enhance the localization of the reconstructed objects by improving the axial (z) resolution and the lateral (x,y) resolution of the reconstructed objects [28-30], and it can also be applied for particle tracking in a particle flow [17, 26, 31-39]. We investigate quantitative 3DD in holography below.

## 6.1 Hologram formation and reconstruction

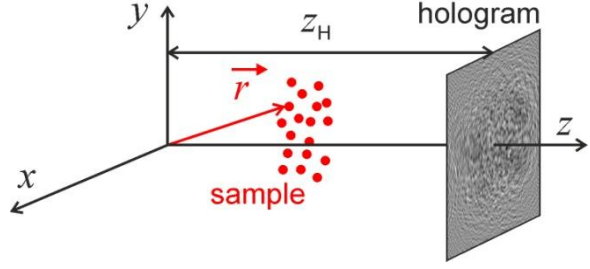

Fig. 7. Illustration of hologram recording and the symbols used.

The hologram recording and reconstruction are illustrated in Fig. 7. A 2D hologram is formed by superposition of the wave diffracted by the sample $u_S(\vec{r})$ and a reference wave $u_R(\vec{r})$ at some distance $z_H$: $\tilde{H}(\vec{r}) = |\tilde{u}_S(\vec{r}) + \tilde{u}_R(\vec{r})|^2$, where the tilde denotes the two-dimensionality, i.e., the function is defined only in the detector's plane, as shown in Fig. 7. After subtraction of the constant background $|\tilde{u}_R(\vec{r})|^2$ and in the approximation of a weak object wave $|\tilde{u}_S(\vec{r})|^2 << |\tilde{u}_R(\vec{r})|^2$, the distribution of the hologram becomes: $\tilde{H}(\vec{r}) \sim \tilde{u}_S(\vec{r})\tilde{u}_R^*(\vec{r}) + \tilde{u}_S^*(\vec{r})\tilde{u}_R(\vec{r})$,

where the first term is the object (sample) wave term and the second term is the term describing its twin image. The reconstruction process consists of illumination of the hologram with the reference wave and backward propagation to the object position. In the first step, the multiplication of the hologram with the simulated complex reference wave $\tilde{u}_R(\vec{r})$ provides the fully restored 2D complex-valued distribution of the object wave in the hologram plane: $\tilde{u}_R(\vec{r})\tilde{H}(\vec{r}) \propto \tilde{u}_S(\vec{r})$. In the second step, the obtained 2D complex-valued wavefront is propagated backward from the hologram plane to the sample's location, which can be calculated as mentioned above, by employing the Huygens-Fresnel diffraction integral, Rayleigh-Sommerfeld diffraction, ASM or other wavefront propagation methods. The reconstructed distribution $u_S^{(rec)}(\vec{r})$ is not the distribution of the sample $s(\vec{r})$ but approximately the distribution of the wavefront diffracted by the sample, as illustrated in Figs. 8(a) and (b). At any given 2D plane, the reconstructed wavefront $u_S^{(rec)}(\vec{r})$ displays the in-focus sample parts but also contains the out-of-focus signal from the sample parts that are located in other planes, which contaminate the reconstruction, as illustrated in Fig. 8(b). However, $u_S^{(rec)}(\vec{r})$ can be assumed to approximately describe the distribution of the diffracted wavefront. The PSF $p(\vec{r})$ is created as a 3D wavefront reconstructed from hologram of a point-like source whose position is approximately in the center of the sample distribution, Fig. 8(c)-(d). 3DD of $u_S^{(rec)}(\vec{r})$ is then performed with $p(\vec{r})$, and the result is the recovered sample distribution $s(\vec{r})$, Fig. 8(e)-(g).

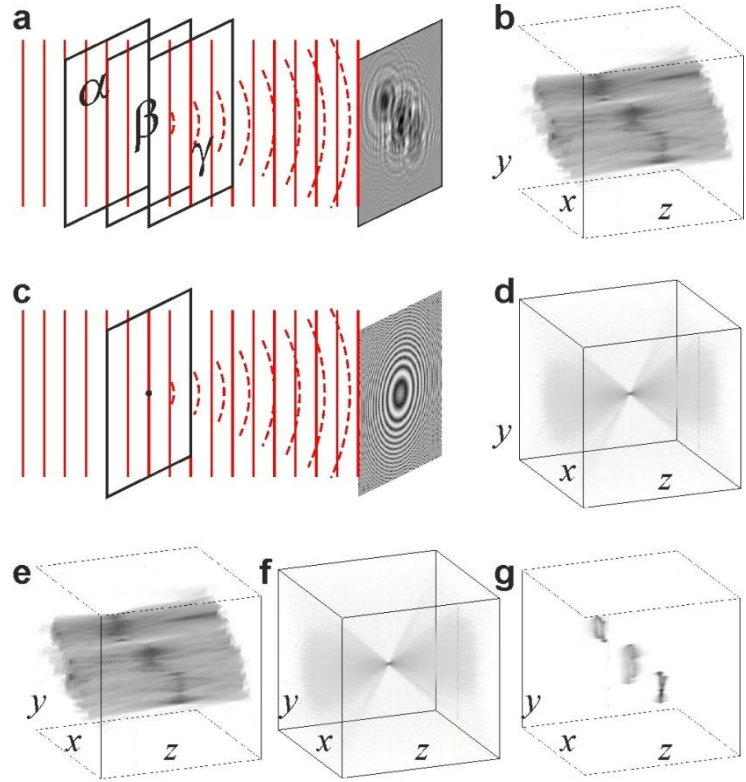

Fig. 8. Principle of 3DD in digital holography. (a) Recording of a hologram of a 3D sample. (b) 3D distribution of the amplitude of the complex-valued wavefront reconstructed from the hologram shown in (a). (c) Recording of the hologram of a point absorber. (d) 3D distribution of the amplitude of the complex-valued 3D point spread function (3D PSF) obtained by reconstruction of the hologram shown in (c). (e)–(g) 3DD of 3D reconstructed wavefront (e) with the 3D PSF (f) gives twin image and out-of-focus signal-free reconstruction of 3D sample (g).

## 6.2 PSF in holography

For off-axis holography, 3DD can be applied to the reconstructed wavefront as described above and quantitatively correct 3D sample distributions can be recovered. For in-line holography, where the reference and object waves share the same optical axis, the task of extracting the sample distribution and obtaining its quantitatively correct reconstruction is somewhat trickier. In in-line holography, the 3D wavefront diffracted by a sample can be written as $u(\vec{r}) = t(\vec{r}) \otimes p(\vec{r})$, where $t(\vec{r}) = \exp[-a(\vec{r})]\exp[i\varphi(\vec{r})]$ is the transmission function distribution, and $a(\vec{r})$ and $\varphi(\vec{r})$ are the absorption and phase distributions, respectively. The transmission function can be re-written as $t(\vec{r}) = 1 + o(\vec{r})$, where $o(\vec{r}) = t(\vec{r}) - 1$, which gives $u(\vec{r}) = t(\vec{r}) \otimes p(\vec{r}) = 1 + o(\vec{r}) \otimes p(\vec{r})$. In this representation, term 1 describes the reference wave and the term $o(\vec{r}) \otimes p(\vec{r})$ describes the distribution due to diffraction on the sample, this distribution is reconstructed from the hologram. $o(\vec{r})$ is then recovered from $o(\vec{r}) \otimes p(\vec{r})$ by applying 3DD. For both off-axis and in-line holography, $p(\vec{r})$ can be calculated numerically. The coordinates of the point scatterer should be selected approximately in the center of the sample distribution. The 2D complex-valued distribution of the wavefront from the point scatterer is calculated in the detector plane by using Eqs. (4) - (6). The obtained 2D wavefront is then numerically propagated backward through the sample planes, and thus the 3D PSF $p(\vec{r})$ is obtained.

## 6.3. Examples of real-valued objects

### 6.3.1 Patch

A real-valued sample was chosen in the form of a square patch, with the same parameters as the patch examples above. The parameters of the simulation and reconstruction are provided in Appendix A. 3DD was performed as described above, using Filters2 with a loose support mask of 40 × 40 × 4 pixels, and using Filters3 with a tight mask object support of 11 × 11 × 1.6 pixels. The mismatch between the original $s(i,j)$ and the retrieved $s'(i,j)$ distributions was calculated as:

$$\text{Error} = \sqrt{\frac{1}{N_0^2} \sum_{i,j=1}^{N_0} \left| s(i,j) - s'(i,j) \right|^2}, \quad (13)$$

where $i, j$ are the pixel numbers, $N_0 \times N_0$ is the number of pixels of the patch and the summation is performed only over the patch area.

Two patch samples were simulated. Patch1 was an opaque patch and the transmission function in the sample plane was $t(x,y) = 0$ at the patch location and $t(x,y) = 1$ elsewhere. The amplitude of the wavefront reconstructed from the hologram is shown in Figs. 9(a)–(c). The sample distribution obtained after 3DD using Filters2 for 9999 iterations is shown in Figs. 9(d)–(f). The results for Filters3 after 1000 iterations are shown in Figs. 9(g)–(i). Both retrieved distributions show correct spatial distribution of the sample. The retrieved transmission values are more correct in the case of 3DD with Filters3. The error calculated by Eq. (13) is 0.27 for Filters2 (loose support) and 0.17 for Filters3 (tight mask object support).

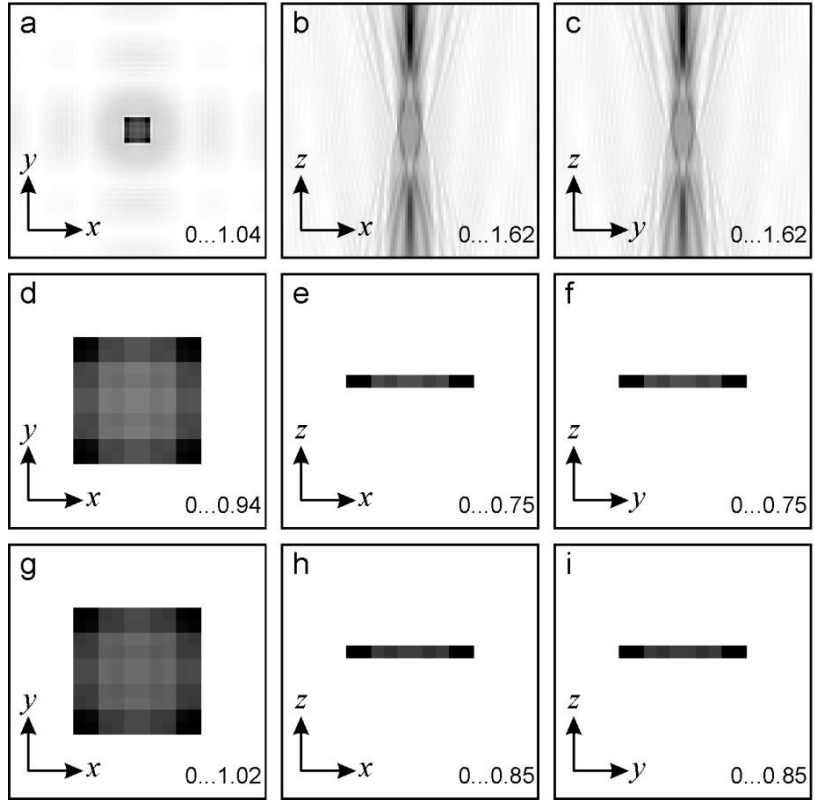

Fig. 9. 3DD of the complex-valued wavefront reconstructed from a hologram of an opaque patch. 2D amplitude distributions in the $(x, y, 0)$, $(x, 0, z)$ and $(0, y, z)$ planes are shown. The contrast is inverted. (a)–(c) Amplitude distributions of the reconstructed wavefront. (d)–(f) Amplitude distribution of the sample distribution obtained after 3DD with Filters2 (loose sample support) and (g)–(i) after 3DD with Filters3 (tight mask object support). In (d)–(i), only the central 20 × 20 pixel regions are shown. The values given at the bottom are the range of the amplitude values.

In the next example, a partially transparent patch (Patch2) was investigated to mimic more realistic samples that exhibit finite absorption (transmission ≠ 0) rather than full absorption (transmission = 0). The transmission function of Patch2 was $t(x,y) = 0.2$ at the patch location and $t(x,y) = 1$ elsewhere. The amplitude of the wavefront reconstructed from the hologram is shown in Figs. 10(a)–(c). The sample distribution obtained after 3DD using Filters2 for 9999 iterations is shown in Figs. 10(d)–(f) and using Filters3 after 1000 iterations is shown in Figs. 10(g)–(i). Also here, both retrieved distributions show the correct spatial distribution of the sample. The retrieved amplitude values are more correct in the case of 3DD with Filters3. The error calculated by Eq. (13) is 0.19 for Filters2 (loose support) and 0.08 for Filters3 (tight mask object support).

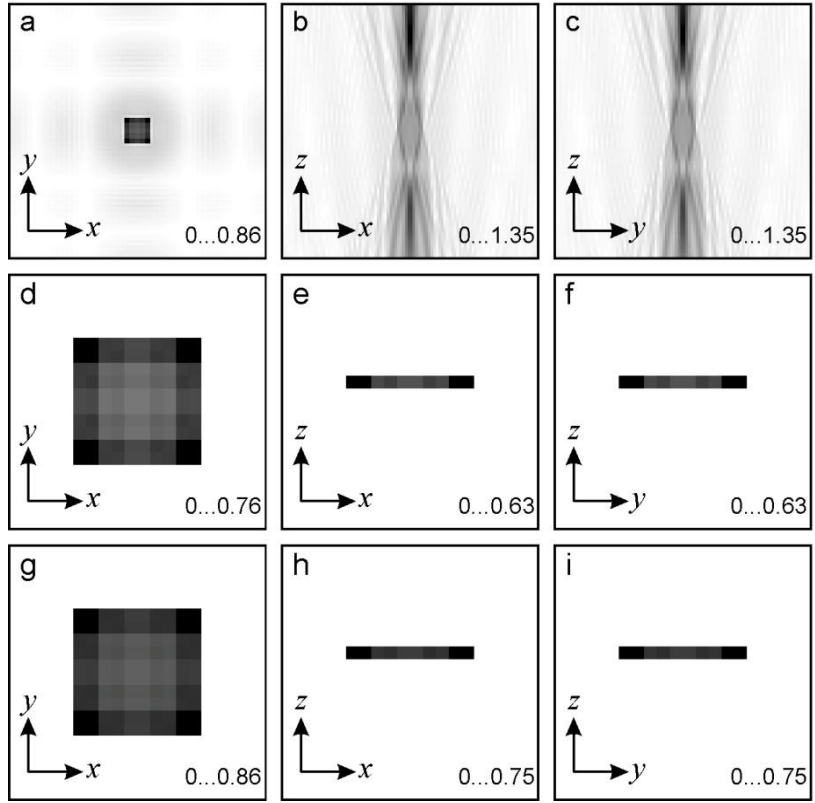

Fig. 10. 3DD of wavefront reconstructed from a hologram of a partially transparent patch. 2D amplitude distributions in the $(x,y,0)$, $(x,0,z)$ and $(0,y,z)$ planes are shown. The contrast is inverted. (a)–(c) Amplitude distributions of the reconstructed wavefront. (d)–(f) Amplitude distribution of the sample distribution obtained after 3DD with Filters2 (loose sample support) and (g)–(i) after 3DD with Filters3 (tight mask object support). In (d)–(i), only the central 20 × 20 pixel regions are shown. The values given at the bottom are the range of the amplitude values.

### 6.3.2 α, β and γ sample

Another 3D real-valued sample was chosen in the form of "α", "β" and "γ" letters positioned at different z planes. The hologram was simulated and reconstructed with the parameters provided in Appendix A. The distance z1-to-detector was 89.6 mm ("α"), z2-to-detector was 80 mm ("β") and z3-to-detector was 70.4 mm ("γ"). The amplitude distributions obtained by conventional reconstruction are shown in Fig. 11(a)-(c). 3DD was performed as described above using Filters2 and Filters3, the corresponding results are shown in Fig. 11(h) and Fig. 11(d)-(f),(i), respectively. The loose binary mask in Filters2 was created by (i) calculating the 3D convolution of the sample distribution with the low-pass filter given by Eq. (12); (ii) normalization of the resulting distribution by division with its maximal value; (iii) and by setting the values that exceed 0.3 to 1, and otherwise to zero. In addition, a positive absorption filter was applied by setting the amplitude of the sample transmission function to 1 where it exceeded 1 [24]. The results of 3DD shown in Fig. 11 demonstrate that after 3DD, the sample distribution is free from the out-of-focus and twin image signals, and is quantitatively correctly retrieved.

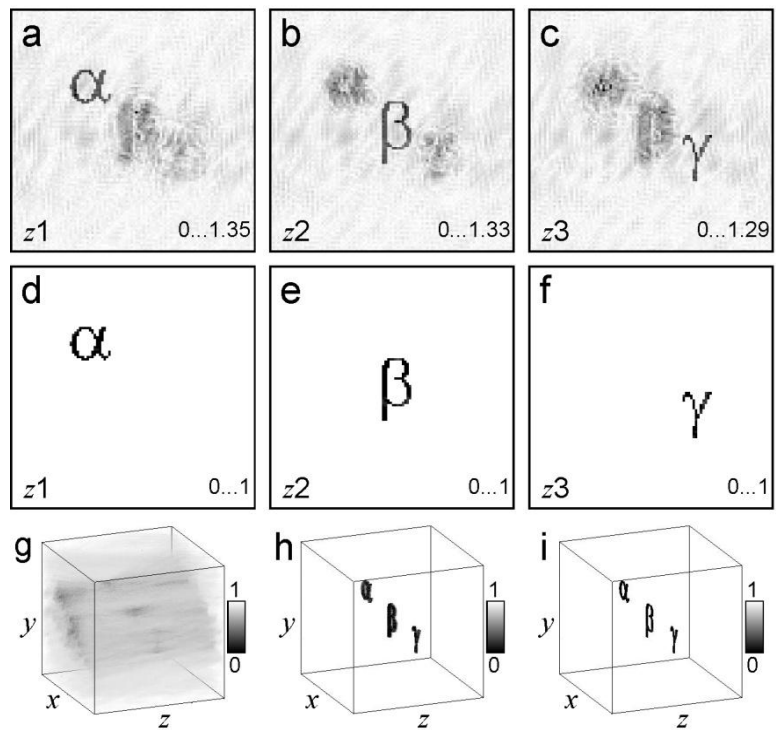

Fig. 11. 3DD of wavefront reconstructed from a hologram of a 3D sample consisting of "$\alpha$", "$\beta$" and "$\gamma$" letters. The contrast is inverted. (a)–(c) 2D amplitude distribution of the reconstructed 3D wavefront in the z1, z2 and z3 planes. (d)–(f) Amplitude of the sample distribution obtained after 3DD with Filters3 (tight mask object support) in the same planes as in (a)–(c). (g) 3D distribution of the amplitude of the reconstructed wavefront. 3D distribution of the sample amplitude obtained after 3DD with Filters2 (loose sample support) (h) and Filters3 (tight mask object support) (i). The values given at the bottom (a) - (f) are the range of the amplitude values.

A recommended protocol of 3D sample retrieval from its 3D diffracted wavefront includes the following steps: (1) 3DD without mask support, as demonstrated in [5] to obtain an initial distribution, (2) 3DD with loose mask support to refine the object spatial distribution and finally (3) 3D deconvolution with a tight mask object support.

## 6.4 Note on phase objects

In in-line holography, phase objects cannot be reconstructed from their holograms by simple wavefront propagation [40] and an iterative reconstruction routine must be applied [40, 41]. Because the reconstructed 3D wavefront is obtained by simple wavefront propagation, it does not contain accurate phase information. However, this 3D wavefront is the input signal for 3DD. Thus, accurate phase information is already lost at this stage and phase objects cannot be restored artefact-free and quantitatively accurate from their reconstructed wavefronts by 3DD. The situation is different for off-axis holography, where the reconstructed wavefront contains correct phase information. In this case, 3DD can be applied, as demonstrated above for 3DD of complex-valued wavefronts.

## 7. Resolution enhancement

3DD improves the resolution of the retrieved 3D samples. To quantitatively estimate the resolution improvement, we recall the classical lateral and axial resolution criteria [42]:

$$R_{\text{Lateral}} = \frac{\lambda}{2\text{NA}} \tag{14}$$

and

$$R_{\text{Axial}} = \frac{2\lambda}{\text{NA}^2}, \tag{15}$$

respectively, where NA is the numerical aperture, defined by the effective size of the formed interference pattern, which in general can be different from the detector's size. If the period of fringes is less than double the pixel size, these fringes cannot be resolved and they do not contribute to the interference pattern.

## 7.1 Lateral (*x*, *y*) resolution

The lateral resolution given by the Abbe criterion (Eq. (14)) can be rewritten as:

$$R_{\text{Lateral}} = \frac{\lambda}{2NA} \approx \frac{z\lambda}{s_D}, \tag{16}$$

where $s_D \times s_D$ is the detector size. From comparing Eqs. (11) and (16), it follows that the pixel size that allows for correct sampling should be equal or larger than the Abbe resolution limit. Thus, one cannot choose a smaller pixel size to increase the lateral resolution because it will cause incorrect sampling and introduce artefacts. In addition, it should be noted that for digitally sampled signals, two point-like objects can be resolved only if they are separated by at least one pixel, which gives the digital resolution limit:

$$R_{\text{Digital}} = 2\Delta. \qquad (17)$$

where $\Delta$ is the pixel size. Thus, the resolution limit is defined by the larger of $R_{\text{Lateral}}$ or $R_{\text{Digital}}$. Due to the correct sampling issue, 3DD can improve the lateral resolution to the resolution limit but not beyond. This is illustrated in Fig. 12, where 3DD is demonstrated for an experimental hologram of microspheres (4 μm in diameter) on glass acquired with laser light at 532 nm and a distance 970 μm with the hologram size of 550 μm × 550 μm, which gives the resolution limit according to Eq. (16) of $R_{\text{Lateral}} = 1.01$ μm. When such a hologram is sampled with 1000 × 1000 pixels, the pixel size is $\Delta = 0.55$ μm. Two selected spheres at 5.5 μm (10 pixels) apart can be clearly resolved, as shown in Fig. 12(a). When the same hologram is sampled with 200 × 200 pixels (low-resolution hologram), the pixel size increases to $\Delta = 2.75\,\mu m$, but the resolution limit according to Eq. (16) remains the same. The digital resolution limit (Eq. (17)) becomes $R_{\text{Digital}} = 2\Delta = 5.5$ μm, which thus defines the resolution limit. The two spheres become almost unresolved, as shown in Fig. 12(b). After applying 3DD, the lateral resolution is greatly improved, reaching the digital resolution limit and the two spheres become clearly resolved, as shown in Fig. 12(c).

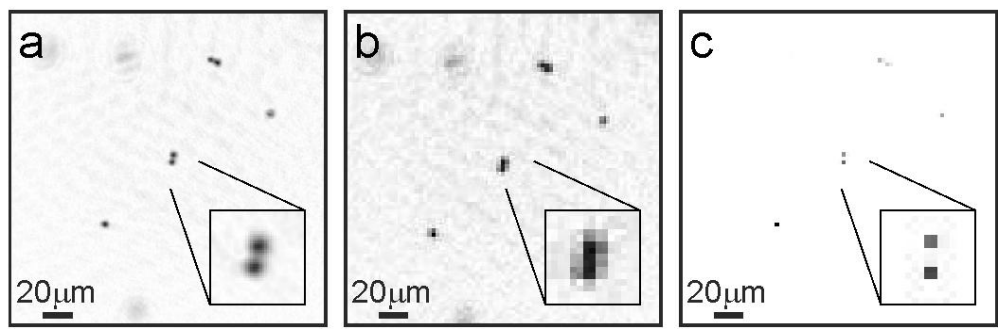

Fig. 12. Enhancement of lateral resolution after 3DD. Amplitude distribution reconstructed from (a) high- and (b) low-resolution holograms. (c) Amplitude distribution of the sample after 3DD. The contrast is inverted.

## 7.2 Axial (z) resolution

The axial resolution can be improved beyond the classical limit by 3DD, owing to the possibility of arbitrarily choosing the z-step size in the reconstruction. An example is shown in Fig. 13, where two point-like sources at different z locations are simulated using the parameters in Appendix A. At these parameters, the axial resolution limit according to Eq. (15) is 6.4 mm. When the distance between the point-like sources is 6.4 mm, the amplitude of the coherent sum of the two wavefronts show two maxima; however, not at the correct positions, as shown in Figs. 13(a) and (b). This effect has been previously discussed in ref. [42]. The two point-like sources appear as sharp localized peaks at the correct positions only after applying 3DD, as illustrated in Figs. 13(c) and (d). By performing numerical simulations, it was established that the minimal z distance between the sources when they still can be resolved after 3DD is 1.6 mm, which is four times smaller than the axial resolution given by the classical criterion in Eq. (15), as shown in Fig. 13(e)-(h). Iterative 3DD in these examples was applied using Gold's algorithm without normalization (Eq. (9)), with Wiener division in step (ii), $\beta_0$ = 0.01, without a mask object support, for 10000 iterations. The resolution enhancement was also checked against the noise. Gaussian-distributed noise was added to the 2D complex-valued wavefront distribution in the detector plane, so that the corresponding intensity distribution exhibited a signal-to-noise ratio (SNR) of 5. The 3D wavefront was obtained by propagating the noisy 2D wavefront distribution from the detector plane backward as explained in Appendix A. 3DD was then applied to the obtained 3D wavefront distribution. The two point-like sources appear as clearly resolved as in the noise-free case, as shown by the red curve in Fig. 13(h). The reason why 3DD appears to be so independent of the SNR in this case can be explained as follows. For a point source, its intensity distribution in the detector plane is nearly a constant distribution, and an added noise can result in a low SNR distribution. However, when the wavefront is propagated backward it converges to an intense peak at the source origin. The intensity near the source location is much higher and exhibits a higher SNR than that of the surrounding background, and it also dominates the outcome of 3DD. It should also be noted that SNR can typically be improved by using a longer acquisition time during the experiment.

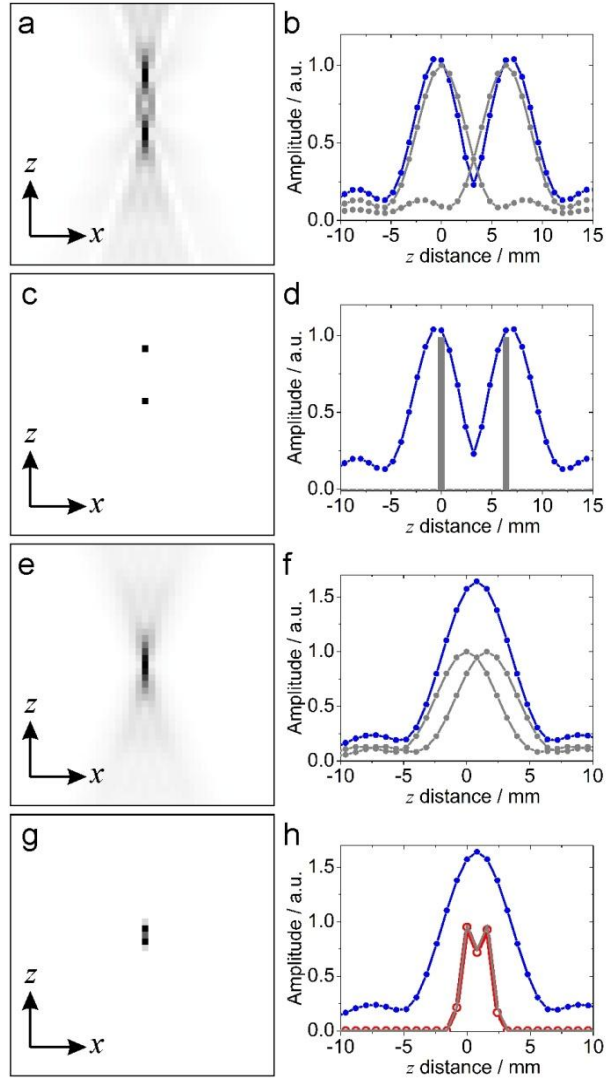

Fig. 13. Axial resolution enhancement by 3DD. Two point sources are z separated by (a)–(d) 6.4 mm and (e)–(h) 1.6 mm. (a) and (e) 2D amplitude distributions in the (x, 0, z) plane before and (c) and (g) after 3DD; only the central regions of 40 × 40 pixels are shown. The contrast is inverted. (b) and (f) Amplitude profiles along the z-axis before and (d) and (h) after 3DD. The gray curves show the amplitudes of individual point-like sources and the blue curves show the amplitude of the coherent sum of the wavefronts. The red curve in (h) shows the amplitude profile after 3DD performed on 3D wavefront obtained from a noisy 2D complex-valued distribution (SNR = 5), as explained in the text.

## 8. Discussion and outlook

In conclusion, we studied 3DD applied to complex-valued wavefronts diffracted by a 3D sample. Such wavefronts can be directly measured by employing an interferometric technique, such as off-axis holography. In the case of in-line holography, the reconstructed wavefront contains not only the wavefront diffracted by the sample but also artifactual signals from the twin image and the object terms. These artifactual signals are removed after 3DD is applied.

Samples consisting of point-like objects can be retrieved from their diffracted wavefronts by non-iterative 3DD. Amplitude and phase of continuous, extended samples, including complex-valued (phase) samples, can be quantitatively correctly retrieved only by applying an iterative 3DD.

Two algorithms for iterative deconvolution were presented, namely, Gold's algorithm and the Richardson-Lucy (RL) algorithm. It was observed that when applied to optical fields, the RL algorithm quickly diverges while Gold's algorithm demonstrates stable convergence and correct results.

The most effective protocols for quantitatively correct sample restoration were presented. A recommended protocol of 3D sample retrieval from its 3D diffracted wavefront includes the following steps: (1) 3DD without mask support as demonstrated in [5] to obtain an initial distribution, (2) 3DD with a loose mask support to refine the object spatial distribution, and finally (3) 3D deconvolution with a tight mask object support. This approach does not require any a priori information regarding the sample.

3DD can greatly improve the resolution. The lateral resolution can be enhanced to the resolution limit but not beyond due to the sampling issue. The axial resolution can be improved to be at least four times better than the resolution limit, owing to the possibility of arbitrarily choosing the z-step size in the reconstruction.

## Appendix A: Simulation parameters

For the simulations shown in this study, the parameters are selected to ensure correct sampling and are wavelength $\lambda$ = 500 nm, sample and detector area sizes $s_D \times s_D$ = 2 mm × 2 mm and sampling with 100 × 100 × 100 pixels. The simulation of the scattered wavefront is carried out as illustrated in Fig. 4(a) and (b). The wavefront diffracted from the patch is calculated at a distance $z$ = 80 mm from the patch and then propagated backward for the range of distances from $z$ = 40 mm to $z$ = -40 mm (or from z = 40 to 120 mm in negative z direction when counted from the detector plane) at a z-step of 0.8 mm with 100 steps in total.

# MATLAB codes

## 3D non-iterative deconvolution

https://ch.mathworks.com/matlabcentral/fileexchange/59984-3d-deconvolution-volumetric-deconvolution-of-wavefronts

## 3D iterative deconvolution

https://ch.mathworks.com/matlabcentral/fileexchange/85458-3d-deconvolution-iterative